# Current-Excited Magnetization Dynamics in Narrow Ferromagnetic Wires

Yoshihiko Togawa[1]*, Takashi Kimura[1,2], Ken Harada[1,3], Tetsuya Akashi[3,4], Tsuyoshi Matsuda[1,3], Akira Tonomura[1,3] and Yoshichika Otani[1,2]

[1]*Frontier Research System, the Institute of Physical and Chemical Research (RIKEN), Hatoyam, Saitama, 350-0395, Japan /Hirosawa, Wako, Saitama 351-0198, Japan*
[2]*Institute for Solid State Physics, The University of Tokyo, Kashiwanoha, Kashiwa, Chiba 277-8581, Japan*
[3]*Advanced Research Laboratory, Hitachi, Ltd., Akanuma, Hatoyama, Saitama 350-0395, Japan*
[4]*Hitachi Instruments Service Co., Ltd., Yotsuya, Shinjuku-ku, Tokyo 160-0004, Japan*

**We investigate the current-excited magnetization dynamics in a narrow ferromagnetic Permalloy wire by means of Lorentz microscopy, together with the results of simultaneous transport measurements. A detailed structural evolution of the magnetization is presented as a function of the applied current density. Local structural deformation, bidirectional displacement, and magnetization reversal are found below the Curie temperature with increasing the current density. We discuss probable mechanisms of observed features of the current-excited magnetization dynamics.**

*All correspondence should be addressed to Y. T. (ytogawa@harl.hitachi.co.jp)

When the spin-polarized current is applied to a narrow ferromagnetic wire containing a magnetic domain wall, the conduction electron spins exert the torque on the constituent localized magnetic moments in the wall. The induced spin torque rotates successively the localized magnetic moments and thereby the magnetic domain wall is driven along the spin current. The responsible mechanism of this domain wall dynamics is based on the spin (angular momentum) transfer.[1-4]

Recent experiments on the pulsed current-driven domain wall motion[5-11] proved that the spin transfer is responsible for the domain wall dynamics in the magnetic narrow wire. However, imaging techniques revealed rather complicated dynamics of the domain wall. For instance, structural transformation from a vortex to a transverse domain wall impedes its propagation even under the applied current-pulse.[9] In addition, the current pulse induces considerable Joule heating[12] of which thermal excitation may affect the domain wall dynamics. Therefore, systematic investigation on the development of the magnetized state as a function of the driving force is required to clarify detailed behavior of an individual domain wall and the underlying physics of the domain wall dynamics.

In the present study, we employ Lorentz microscopy[13] that enables the real-time observation using video (30 frame/s) in a high spatial resolution (down to 10 nm).[14] Magnetization in the ferromagnetic wire is qualitatively analyzed by Lorentz microscopy. Quantitative distribution of the magnetic flux line is obtained by means of electron holography.[15] While observation, a current pulse with variable amplitude is applied every one second to the observed magnetic wires and the wire resistance is measured to monitor the wire temperature.[12] This experimental scheme saves enormous time to obtain a sequence of snapshots of the magnetization before and after the current-pulse application as compared with the scanning probe microscopy. The device with the magnetic wires is mounted on a special holder for the current application, installed in the column of the 300 kV field-emission transmission electron microscope above the objective lens where the sample is free from the magnetic field generated by the electromagnetic lenses.[13]

The sample is Permalloy zigzag wire with thickness of 30 nm and width of 500 nm. The magnetization with spatial gradient naturally formed at every corner of the zigzag wire favors the spin torque, assuring of the appropriate sample geometry for the present investigation. Since the velocity of the domain wall is found to be about 10 m/s around the threshold current density,[8, 9)] the domain wall moves at most over two or three straight segments of the wire when the pulse with the duration of less than 1 μs is applied. In this case, the domain wall displacement can be traced. As an initial state, head-to-head or tail-to-tail domain walls are generated at every corner of the zigzag wire as shown in Figs. 1(b) and 1(c) by applying an in-plane magnetic field perpendicular to the averaged direction of the wire. Although a vortex domain wall is preferable to the wire with given dimensions,[16)] the magnetization structure formed at the corner during demagnetization process often becomes a distorted transverse domain wall as shown in Figs. 1(c)-1(e).

The wire resistance as a function of the current density at the pulse duration of 200 ns is shown in Fig. 2(a). The resistance gradually increases because of Joule heating, and presents a kink at the beginning of a linear dependence above the current density of $2.30 \times 10^{11}$ A/m$^2$. As the temperature dependence of the resistance is known to exhibit a kink at the Curie temperature $T_C$ in ferromagnetic materials,[17)] the kink in Fig. 2(a) corresponds to $T_C$ since the temperature monotonously increases with increasing the current density. Characteristic changes in the magnetic states are observed below $T_C$ as the current density increases. The magnetization dynamics induced by the application of 200 ns current pulses is classified into three regions; (1) local structural deformation including a minute oscillation above $1.84 \times 10^{11}$ A/m$^2$, (2) bidirectional domain wall displacement above $2.09 \times 10^{11}$ A/m$^2$, and (3) magnetization reversal accompanying nucleation and annihilation of paired domain walls above $2.14 \times 10^{11}$ A/m$^2$. The magnetic state develops similarly below $T_C$ when the pulse duration is set between 100 ns and 1 μs. Note that similar tendencies are observed for five samples investigated in this study.[14)]

First, in the region (1), the magnetic configuration in the wall deformed locally by the applied current pulse relaxes into the initial structure before the following pulse application. This oscillatory behavior is observed until the more stable configuration is formed by the current pulse application. Indeed, the transverse domain wall at the corner in Fig. 2(b) transforms to the vortex domain wall in Fig. 2(c) through this process.[14)] Single vortex domain wall at the straight segment of the wire often shows a minute oscillation of the vortex core with the amplitude of typically about 50 nm.

With increasing the current density in the region (2), the domain wall is displaced at the threshold current $J_{th}$ defined as the current where the domain wall travels over the distance of the wire width. Figures 2(c) and 2(d) show that the vortex domain wall at the corner is displaced into the straight segment of the wire. Depending on positions, the displaced domain wall continues to move or stops by the subsequent application of the current pulse. Once the domain wall is pinned, the current pulse with the amplitude larger than $J_{th}$ is required to depin. It should be noted that the displacement occurs both along and against the electron flow as shown in Figs. 2(c)-2(g).[14] In other words, the magnetic domain expands and shrinks in the vicinity of the original position. The displaced domain walls change their magnetic configurations characterized by the number of vortices, their chiralities, sequence, and positions. Domain walls in Figs. 2(e)-2(g) consist of up to three chained vortices. More than ten vortices are chained in one straight segment of the wire at higher current densities. When a vortex at the end of chained vortices is strongly pinned, the rest of vortices oscillate as if they were connected to each other by a spring.

Similar displacement occurs frequently with further increasing the current density. However, characteristic finding in the region (3) is that the displacement of adjacent domain walls leads to the magnetization reversal as shown in Figs. 2(h) and 2(i). Under the electron flow toward the left, the domain wall on the left hops to the right toward the other domain wall. Two domain walls eventually merge and annihilate, and turn back to the uniformly magnetized state of the wire.[14] The magnetization reversal also occurs by forming chained vortices over the domain, connecting adjacent domain walls, and annihilating them. In addition, paired domain walls are nucleated in the domain, which also reverses the magnetization. It should be noted that in most cases, straight segments at both ends of the wire connected to the electrode pads show no distinct change in magnetization while the magnetization is reversed in the wire. This means that additional domain walls are not supplied from the pad but nucleated inside the wire by the applied current. Note that in regions (2) and (3), the magnetic state changes within a frame only when the current pulse is applied.

In the present experiment, local deformation observed at lower current density is consistent with previous studies,[9, 18, 19] indicating that the spin torque works effectively. However, the bidirectional dynamics of the domain wall and the magnetization reversal inside the domain are beyond simple one-directional picture of the domain wall propagation along the current flow.[3] The magnetic field induced by a current (Oersted field) is negligibly small in thin wire and could not explain the bidirectional domain wall displacement.[18] Concerning thermal effect, it is found that $J_{th}$ increases and the wire resistance at the threshold decreases with decreasing the pulse duration as shown in Fig. 2(a). This behavior is reminiscent of thermally-assisted spin torque induced magnetization reversal observed in pillar structures,[20, 21] supporting that the spin torque plays an essential role to drive the domain wall in the present case and that thermal agitation effectively lessens the pinning potential energy barrier for the domain walls. The spin-wave instability followed by the domain wall nucleation[22-25] may explain the bidirectional dynamics of the domain wall and the magnetization reversal inside the domain even under the spin torque. Furthermore, non-adiabatic spin torque[18, 19, 26, 27] should be incorporated. Although the exact mechanism of observed features remains to be clarified, present experimental findings exhibit a wide variety of the magnetization dynamics under the spin torque, and should help further physical understanding of the current-driven magnetization dynamics to enrich the technology of the magnetization manipulation.[28]

We thank Dr. J. Shibata of RIKEN for fruitful discussions and Mr. N. Moriya of Hitachi Ltd. for technical support.

**Figure captions**

Figure 1 (a) Experimental principle underlying Lorentz microscopy. Since incident electrons are deflected by the Lorentz force due to the magnetization inside the sample, excessive and deficient distributions of the electron intensity are produced around an image of the sample in a defocused plane. Thus, bright (dark) linear contrast is seen along the sample, locating on the left (right) of the magnetization direction in the overfocused plane below the sample plane. The bright and dark contrast lines are reversed against the magnetization direction when defocusing in the opposite direction in the underfocused condition. Whole Lorentz images shown in Figs. 1 and 2 and videos on the web[14] are underfocused images. (b) Underfocused Lorentz micrograph of the Permalloy zigzag wire. Head-to-head or tail-to-tail domain walls indicated by arrows are generated as an initial state through demagnetization process. Enlarged image of a part of the wire is shown in (c). Five wires connected to large Permalloy pads are fabricated by lift-off technique using electron-beam lithography on the 50 nm thick $Si_3N_4$ membrane supported by a Si wafer. Resistivity of Permalloy wire is 14.6 μΩ cm at room temperature. One of pads is electrically grounded and positive current pulses are applied every one second to the wire. Whole experiments in this study are done at room temperature. (c, d) Lorentz micrographs of the transverse and the vortex domain wall. The vortex domain walls consist of three chained vortices with alternate chiralities, recognized as bright or dark spots. (e) Twice phase-amplified interference micrograph of the vortex domain wall in (d). The magnetic flux lines lie along the wire in the areas with no domain walls, whereas they circulate around the vortex core and penetrate from the vacuum into the wire around the domain wall.

Figure 2 (a) The wire resistance as a function of the current density flowing in the wire at the pulse duration of 200 ns and 120 ns. For 200 ns current pulses, local structural deformation, domain wall displacement, and magnetization reversal are found at $1.84 \times 10^{11}$ A/m$^2$, $2.09 \times 10^{11}$ A/m$^2$ (threshold current density $J_{th}$, and $2.14 \times 10^{11}$ A/m$^2$, respectively. The broken line on the data is a guide for the eye. The kink of the resistance found at $2.30 \times 10^{11}$ A/m$^2$ and 2350 Ω indicates the Curie temperature $T_C$ for the observed wire. For 120 ns current pulses, domain wall displacement occurs at $J_{th}$ of $3.14 \times 10^{11}$ A/m$^2$ and 1780 Ω. Characteristic micrographs taken at 200 ns current pulses are shown in (b-g). (b) Initially distorted magnetized state at $3.30 \times 10^{10}$ A/m$^2$. (c) Structural transformation into a vortex domain wall at $1.91 \times 10^{11}$ A/m$^2$ and 1750 Ω. (d) Domain wall displacement at $2.09 \times 10^{11}$ A/m$^2$ and 1920 Ω. (e-g) Forward and backward domain wall movements (bidirectional displacement) at $2.17 \times 10^{11}$ A/m$^2$ and 2080 Ω. (h, i) Paired domain walls annihilation (magnetization reversal) under 1 μs current pulses of $9.66 \times 10^{10}$ A/m$^2$ in a different sample. The wire resistances at annihilation and $T_C$ are 2200 Ω and 2450 Ω, respectively. Domain walls are indicated by arrow heads. The direction of the electron flow and the magnetization are schematically indicated by arrows. See also a video on the web.[14]

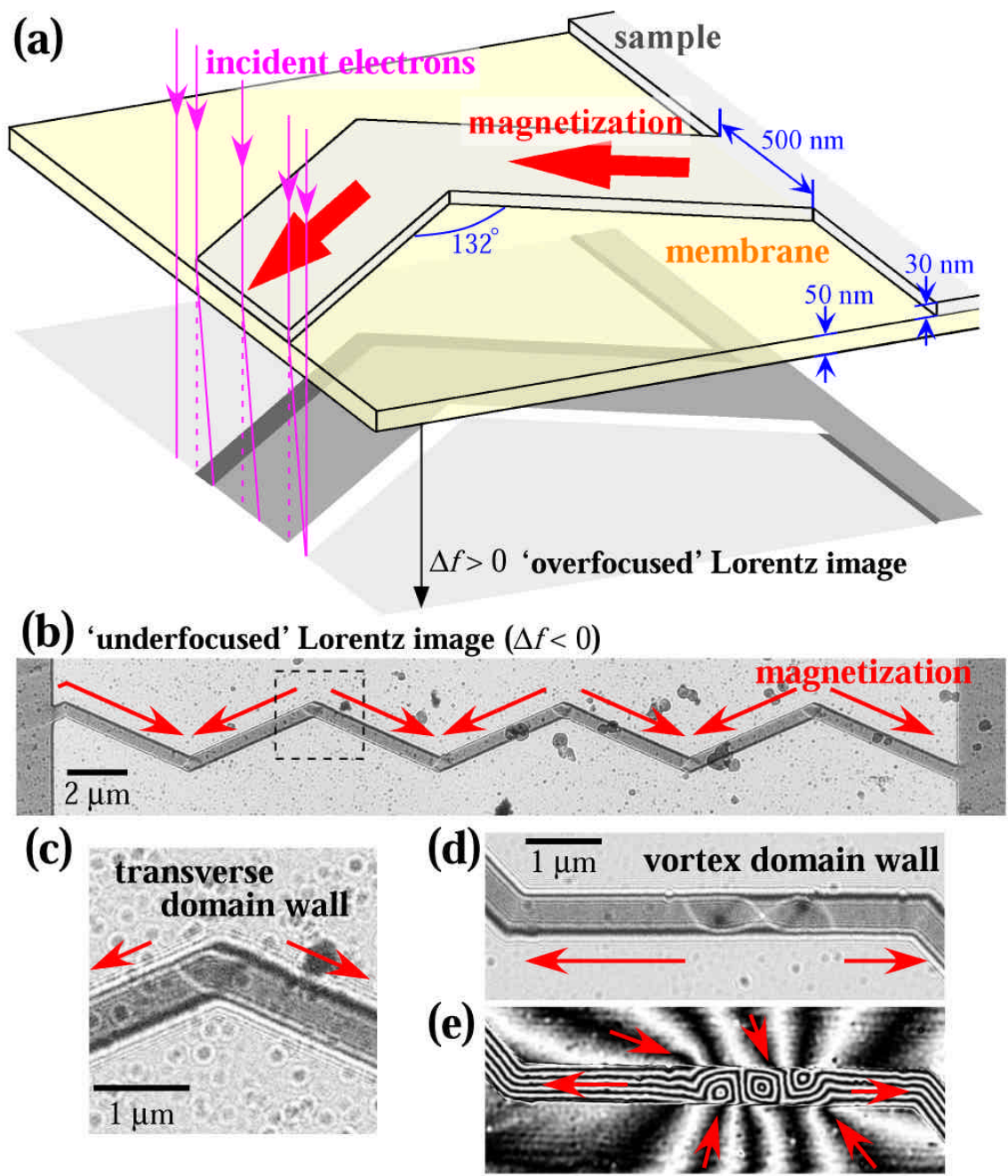

Fig. 1. Y. Togawa *et al*., "Current-Excited Magnetization Dynamics …"

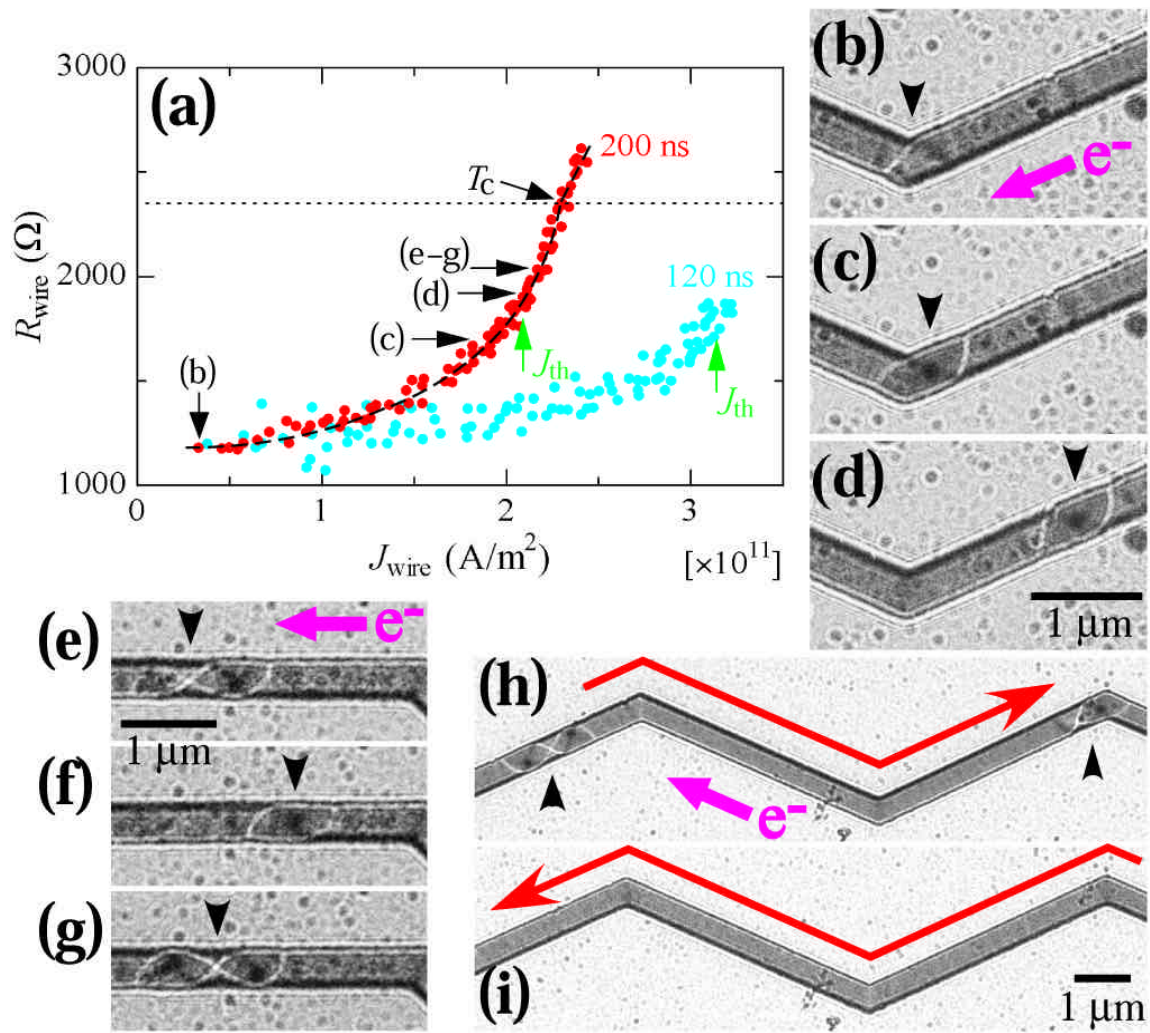

Fig. 2. Y. Togawa *et al.*, "Current-Excited Magnetization Dynamics …"